\documentclass{ws-ijmpa}

\newcommand{\be}{\begin{equation}}
\newcommand{\ee}{\end{equation}}
\newcommand{\bea}{\begin{eqnarray}}
\newcommand{\eea}{\end{eqnarray}}
\newcommand{\bean}{\begin{eqnarray*}}
\newcommand{\eean}{\end{eqnarray*}}

\newcommand{\gapproxeq}{\lower
.7ex\hbox{$\;\stackrel{\textstyle >}{\sim}\;$}}
\newcommand{\lapproxeq}{\lower
.7ex\hbox{$\;\stackrel{\textstyle <}{\sim}\;$}}

\begin{document}

\markboth{Qiang Zhao}
{Restricted locality of quark-hadron duality ...}

%
\catchline{}{}{}{}{}
%

\title{Restricted locality of quark-hadron duality in exclusive meson 
photoproduction reactions above the resonance region
}

\author{\footnotesize QIANG ZHAO$^1$\footnote{
Email address: qiang.zhao@surrey.ac.uk} 
and FRANK E. CLOSE$^2$\footnote{
Email address: F.Close1@physics.ox.ac.uk}}

\address{1) Department of Physics, University of Surrey\\
Guildford, Surrey GU2 7XH, 
The United Kingdom\\
2) Department of Theoretical Physics,
University of Oxford, \\
Keble Rd., Oxford, OX1 3NP, The United Kingdom
}

\maketitle

\pub{Received (Day Month Year)}{Revised (Day Month Year)}

\begin{abstract}

We show how deviations 
from the dimensional scaling laws
for exclusive processes may be related to a breakdown in the locality of 
quark-hadron duality, i.e. the ``restricted locality". 
For exclusive reactions like meson photo- and electroproduction above the 
resonance region,
we explore the effects arising from such a local duality breaking and 
propose that it can be a possible 
source for oscillations about the smooth quark counting
rule predicted by pQCD in 
 the 90-degree differential cross sections.

\end{abstract}


In recent years, the high-precision measurement of the nucleon structure 
functions~\cite{jlab1} at the nucleon resonance region 
with high $Q^2$ gives access to 
a direct test of the Bloom-Gilman duality~\cite{bloom}, 
which empirically connects 
the low-energy resonance phenomena with the high-energy scaling behavior. 
Namely, the electroproduction of $N^*$ resonances at low energies and 
momentum transfers averages smoothly around the scaling curve of the nucleon 
structure function $F_2(W^2, Q^2)$ which is measured at large momentum 
transfers. 
Such a senario reflects the dual charactor of the strong interaction 
at the quark and hadron level, and also 
makes the interplay between the pQCD region and the 
conventional resonance region extremely interesting since special 
manifestation of the QCD dual charactor may show up  
due to the interference between the low-energy 
and high-energy processes~\cite{npqcd-pqcd}. In particular, 
it may provide novel insights into the observed 
deviations~\cite{experiment-data,gao} from the pQCD predictions 
in exclusive meson photoproduction reactions 
above the nucleon resonance region, e.g. the energy dependence of 
data at $\theta_{c.m.}=90^\circ$ which oscillates around the value 
predicted by the pQCD quark-counting rules~\cite{bf-mmt}.

The essential realization of duality was summarized
by Close and Isgur~\cite{ClIs01} as ``How does the square of sum 
becomes the sum of squares?". 
To show this, the model of 
two-body spinless constituents~\cite{IJMV,ClIs01,close-zhao}
serves as the simplest example.
The general form for the transition amplitude 
for $\gamma({\bf k}) \Psi_0\to \Psi_N\to \Psi_0 \gamma ({\bf q})$ can be
expressed as
\bea
\label{non-forward-trans}
M&=& 
\sum_N\langle\Psi_0({\bf P}_f,{\bf r})|[e_1e^{-\frac{i{\bf q}\cdot{\bf r}}{2}}
+e_2e^{\frac{i{\bf q}\cdot{\bf r}}{2}}]|\Psi_N\rangle
\langle \Psi_N|  [e_1e^{\frac{i{\bf k}\cdot{\bf r}}{2}}
+e_2e^{-\frac{i{\bf k}\cdot{\bf r}}{2}}]|\Psi_0({\bf P}_i, {\bf r})\rangle \nonumber\\
&=&
\sum_{N=0}^{\infty} \sum_{L=0(1)}^N 
[(e_1^2+e_2^2)d_{00}^L(\theta) +2e_1e_2d_{00}^L(\pi-\theta)]
C_{NL} {\cal F}_{0N}^{(L)}({\bf q})  
{\cal F}_{N0}^{(L)}({\bf k})   \ ,
\eea
where $\Psi_N$
is the harmonic oscillator wave function with the main quantum number $N$, and 
the imaginary part of Compton scattering as a sum over the intermediate 
resonances thus gives access to the structure function of ``nucleon" $\Psi_0$. 
${\cal F}_{N0}^{(L)}({\bf k})$ 
denotes the transition from the initial to the intermediate 
state, and ${\cal F}_{0N}^{(L)}({\bf q})$ for the 
intermediate decay to the final ground state.

At this stage, it is not important to consider details of the 
$L$-dependent factor $C_{NL}$. 
First note that in this simple model 
all terms of $L=odd$ for a given $N$
are proportional to $\cos\theta$, and hence vanish at $\theta=90^\circ$;
thus
we need consider only the parity-even 
states at $\theta=90^\circ$, 
i.e. $N=0$, 2, 4 ... with $L=N$, $N-2$, $\cdots$ 0 for a given even $N$. 
The scattering amplitude at $90^\circ$ can then be expressed as
\bea
\label{trans-even}
M_{\theta=90^\circ}
&=&e_0^2\left[ C_{00}\left(\frac{kq}{2\beta^2}\right)^0
+\frac{1}{2!}\frac 13(-C_{22}+C_{20})\left(\frac{kq}{2\beta^2}\right)^2
\right. \nonumber\\
&&\left.+\frac{1}{4!}\frac{1}{35}(3C_{44}-10C_{42}+7C_{40})
\left(\frac{kq}{2\beta^2}\right)^4 
+\cdots \right] e^{-({\bf k}^2 + {\bf q}^2)/4\beta^2} \ ,
\eea
where $e_0=e_1+e_2$ is the total charge of ``nucleon" $\Psi_0$.

Several points thus can be learned:

i) At the state-degeneracy limit of high energies, 
all the terms with $N\ne 0$ and $L=0,\cdots, N$ in
Eq.~(\ref{trans-even}) would vanish due to the destructive cancellation.
Only the $C_{00}$ term survives:
\be
\label{degenerate}
M= e_0^2C_{00} e^{-\frac{({\bf k}-{\bf q})^2}{4\beta^2}}\Big|_{\theta=90^\circ}
= e_0^2 C_{00} R(t)\Big|_{\theta=90^\circ} \ ,
\ee
where $R(t)$ is recognised as the elastic form factor for the Compton 
scattering~\cite{close-zhao} or more generally the quark-counting-rule-predicted 
scaling factor~\cite{bf-mmt}. So we see that the smooth
behaviour driven by the elastic form factor, which is the essence of the
counting rules, effectively arises from the {\it s}-channel sum combined with 
the destructive interferences among resonances.

ii) Concerning the $L$-degeneracy breaking effect for any given $N$,
each term of $(N,L)$ corresponds to the excitation 
of an intermediate state with given $N$ and $L$. 
The factor $C_{NL}$, which is essentially related to the 
mass position of each state, should be different for the individual
states. 
This leads to oscillations around the simple result of Eq.~(\ref{degenerate}),
due to different partial waves not cancelling locally.
We shall refer to this as ``restricted locality".

Certainly, the simple model can only illustrate 
such a deviation in a pedagogic way. However, 
a similar phenomenon may have existed in physical processes due to 
the restricted locality of duality above the prominent 
resonance region. Notice that deviations from 
quark counting rules exist in 
certain exclusive reactions~\cite{experiment-data,gao}, e.g. 
the 90$^\circ$ differential cross sections of 
$\gamma p\to \pi^+ n$ at $W\sim 3$ GeV exhibit oscillations around
the scaling curves predicted by the counting rules.
We will show how the restricted locality of duality is naturally a
source of such oscillations.

To generalize the above to the physical exclusive processes,
we adopt effective Lagrangians 
for the constituent-quark-meson and quark-photon couplings,  
which were proposed by Manohar and Georgi~\cite{manohar} and extended 
to pseudoscalar meson photoproduction in Refs.~\cite{Li-etal,zhao-pion}.
Briefly, such a treatment highlights the quark correlations
in the exclusive processes (including the Compton scattering). 
We can thus arrive at a general expression for the transition amplitudes
for the {\it s}- and {\it u}-channels, i.e. the direct 
and the virtual resonance excitations:
\begin{eqnarray}
\label{s-deg}
M^{s+u}_{fi}&=& e^{-\frac{{\bf k}^2+{\bf
q}^2}{6\alpha^2}} 
\left\{ \sum_{n=0}^{\infty}({\cal O}^{cc}_d + (-\frac 12)^n {\cal O}^{cc}_c)
\frac{1}{n!}
\left(\frac{{\bf k}\cdot{\bf q}}{3\alpha^2}\right)^{n}\right.
+\sum_{n=1}^{\infty} ({\cal O}^{ci}_d + (-\frac 12)^n {\cal O}^{ci}_c)\nonumber\\
&\times& \frac{1}{(n-1)!}
\left(\frac{{\bf k}\cdot{\bf q}}{3\alpha^2}\right)^{n-1}
+\left.\sum_{n=2}^{\infty} ({\cal O}^{ii}_d + (-\frac 12)^n {\cal O}^{ii}_c)
\frac{1}{(n-2)!}
\left(\frac{{\bf k}\cdot{\bf q}}{3\alpha^2}\right)^{n-2} \right\}\ ,
\end{eqnarray}
where the multiplets are degenerate in $n$.
The spin structures, charge and isospin operators have been 
subsumed in the symbol ${\cal O}$. 
Terms proportional to $({\bf k}\cdot{\bf q}/3\alpha^2)^n$
denote correlations of c.m. - c.m. motions (superscript $cc$), 
while $({\bf k}\cdot{\bf q}/3\alpha^2)^{n-1}$
and $({\bf k}\cdot{\bf q}/3\alpha^2)^{n-2}$ denote the 
c.m. - internal ($ci$) or internal - internal correlations ($ii$), respectively.
The subscript ``d" (``c") denotes the 
direct (coherent) process 
that the photon and meson couple to the same (different) quarks
in the transition. The coherent process
is suppressed by a factor of $(-1/2)^n$ in comparison with 
the direct one for higher excited states.
Note that the conventional Born terms will contribute
to  different parts: the nucleon pole terms included in the 
{\it s}- and {\it u}-channel, and the possible contact term and 
{\it t}-channel charged meson exchange 
included as part of the background terms
due to gauge invariance.
In the low energy regime, the degeneracy in $n$ must 
break. In the SU(6)$\otimes$O(3) symmetry limit, 
for a given $n$ ($\le 2$), multiplets of 
$L$- and $S$-dependent resonances can be separated in this model.
In Ref.~\cite{zhao-pion},
the calculations were in agreement
with experimental data up to $E_\gamma\approx$ 500 MeV.

The dominant term comes from the correlation of the c.m.- c.m. 
motions at the two vertices ($n=0, 1, \cdots$), while 
terms involving the c.m.- internal, or 
internal - internal motion correlation will be suppressed. 
For example, for $n=0$ only the terms involving the c.m. - c.m. 
correlation contribute. These correlations are essentially
the demonstration of the internal degrees of freedom 
of the nucleon system. 

In the high energy limit where the degeneracy achieves, 
the leading term can be expressed
compactly as:
\begin{equation}
\label{s-asymp}
M^{s+u}_{fi}=
({\cal O}^{cc}_d + {\cal O}^{cc}_c e^{-{\bf k}\cdot{\bf q}/2\alpha^2})
e^{-({\bf k}-{\bf q})^2/6\alpha^2} \rightarrow 
({\cal O}^{cc}_d + {\cal O}^{cc}_c)
e^{-({\bf k}-{\bf q})^2/6\alpha^2}\Big|_{\theta=90^\circ} \ ,
\end{equation}
where similar to Refs.~\cite{IJMV,ClIs01,close-zhao} the scaling 
behavior can be realized at small $|t|$
due to the suppression of $e^{-{\bf k}\cdot{\bf q}/2\alpha^2}$
on the coherent term.
At $\theta=90^\circ$, 
both direct and coherent
process contribute and
operators ${\cal O}^{cc}_d$ and ${\cal O}^{cc}_c$ now
are independent of $n$. 
We conjecture that 
a similar factorization for the exclusive process 
may be more general than this
nonrelativistic pictures, as suggested by the pedagogic model~\cite{close-zhao}.
The form of Eq.~(\ref{s-asymp}) 
then represents the realization of duality, in particular, the emergence 
of the empirical quark counting rules after the sum over degenerate
resonances at high energies.
The exponent factor $e^{-({\bf k}-{\bf q})^2/6\alpha^2}$ is thus 
regarded as the ``typical" scaling law factor.
Numerical results can be found in Ref.~\cite{zhao-close-prl}.

To summarize: we have discussed the relation between resonance 
phenomena and the dimensional scaling laws based on the
quark-hadron duality picture 
at $2\lesssim \sqrt{s} \lesssim 3.5$ GeV. 
In contrast to previous 
models for the deviations from quark counting rules,
here we proposed that non-perturbative resonance excitations
are an important source for 
such deviations. At specific kinematics,
e.g. $\theta=90^\circ$, the oscillatory deviations could be
dominantly produced by resonance excitations with ``restricted 
locality", and this argument is general for photon induced two-body 
reactions on the nucleon. 
Although the formulation is nonrelativistic, 
we find it has been valuable to gain insights into 
the regime between the traditional resonance and partonic regions.
We also suggest a non-trivial $Q^2$ dependence for such oscillations.

This work is supported, in part, by the U.K. EPSRC 
Advanced Fellowship (Grant No. GR/S99433/01) 
and The University of Surrey (Ref. RS14/05),  
and the U.K. PPARC, and the
EU-TMR program ``Eurodice'', HPRN-CT-2002-00311.
 Q.Z. thanks  
IHEP (Beijing) for warm hospitalities.

\end{document}